\documentclass[11pt]{article}

\usepackage{braket}
\usepackage{xspace}
\usepackage{booktabs,multirow}
\usepackage{authblk}
  
\usepackage{geometry}
\newgeometry{vmargin={25mm,25mm}, hmargin={27mm,27mm}}

\usepackage{bm}
\usepackage{charter}
\usepackage[bitstream-charter]{mathdesign}

\usepackage{xcolor}
\definecolor{myBlue}{rgb}{0,0.5,0.8}
\usepackage[colorlinks = true,
            linkcolor = myBlue,
            urlcolor  = myBlue,
            citecolor = myBlue,
            anchorcolor = blue]{hyperref}
                                                                                
\usepackage[citestyle=phys,natbib=true]{biblatex}
\bibliography{references.bib}

\usepackage{amsmath}

\usepackage{pgfplots}
\pgfplotsset{compat=newest}
\usepgfplotslibrary{groupplots}
\usepgfplotslibrary{polar}
\usepgfplotslibrary{smithchart}
\usepgfplotslibrary{statistics}
\usepgfplotslibrary{dateplot}
\usepgfplotslibrary{ternary}
\usetikzlibrary{arrows.meta}
\usetikzlibrary{backgrounds}
\usepgfplotslibrary{patchplots}
\usepgfplotslibrary{fillbetween}
\pgfplotsset{%
layers/standard/.define layer set={%
    background,axis background,axis grid,axis ticks,axis lines,axis tick labels,pre main,main,axis descriptions,axis foreground%
}{grid style= {/pgfplots/on layer=axis grid},%
    tick style= {/pgfplots/on layer=axis ticks},%
    axis line style= {/pgfplots/on layer=axis lines},%
    label style= {/pgfplots/on layer=axis descriptions},%
    legend style= {/pgfplots/on layer=axis descriptions},%
    title style= {/pgfplots/on layer=axis descriptions},%
    colorbar style= {/pgfplots/on layer=axis descriptions},%
    ticklabel style= {/pgfplots/on layer=axis tick labels},%
    axis background@ style={/pgfplots/on layer=axis background},%
    3d box foreground style={/pgfplots/on layer=axis foreground},%
    },
}

\newcommand{\norm}[1]{\left\lVert#1\right\rVert}
\newcommand{\ux}{\partial_x u}
\newcommand{\ut}{\partial_t u}
\newcommand{\uxx}{\partial^2_x u}
\newcommand{\uh}{\hat u}

\newcommand{\uht}{\partial_t\uh}
\newcommand{\fm}{\hat{\mathcal U}_\varphi}
\newcommand{\vc}{\hat{\mathcal U}_\theta}
\newcommand{\cost}{\hat{\mathcal C}}
\newcommand{\qmd}{QMoD\xspace}

\newcommand{\caleft}[1]{\multicolumn{1}{l}{#1}}

\colorlet{fc}{gray}
\definecolor{c1}{rgb}{0.992,0.682,0.38}
\definecolor{c2}{rgb}{0.671,0.867,0.643}
\newsavebox\library\sbox\library{
\begin{tikzpicture}[
    hbox/.style = {
      fill = fc, draw=white,
      rounded corners=3.6mm,
      line width = 3.0mm,
      text = fc
    },
  ]

  \node (rhs) at (0.85,2.0) {$\text{LHS}(u)=\bm w \cdot \bm\varphi(u)$};

  \node[hbox] (lib) at (0,0) {$\begin{matrix} w_{12} \\ w_2 \\ w_3 \\ \cdot \\ \cdot \\ \cdot \end{matrix}$};
  \node (lib) at (0,0) {$\begin{bmatrix} w_1 \\ w_2 \\ w_3 \\ \cdot \\ \cdot \\ \cdot \end{bmatrix}$};

  \node (cdot) at (0.70,0) {$\times$};

  \node[hbox] (lib) at (1.5,0) {$\begin{matrix} u \\ \partial u^2_x \\ u\partial u_{xi} \\ \cdot \\ \cdot \\ \cdot \end{matrix}$};
  \node (lib) at (1.5,0) {$\begin{bmatrix} u \\ \partial u^2_x \\ u\partial u_{x} \\ \cdot \\ \cdot \\ \cdot \end{bmatrix}$};

\end{tikzpicture}

}
\newsavebox\ufabox\sbox\ufabox{
\begin{tikzpicture}[
    block/.style = {
      shape = rectangle,
      rounded corners,
      draw, thick,
      minimum height = 2.1cm,
      minimum width  = 3.5cm,
      fill = c2,
    }
    ]

  \foreach \i in {-2,-1,...,2}{
    \def\dx{0.29}
    \draw[thick] (-1.9,\i*\dx) -- (0,\i*\dx);
  }
  \foreach \i in {-2,-1,...,2}{
    \def\dx{0.29}
    \draw[thick] (1.9,\i*\dx) -- (0,\i*\dx);
  }

  \node[block,align=center] (ufa) at (0,0)
    {$\hat u(x,t,\bm\theta) = \braket{\hat{\mathcal C}(x,t)}$};

\end{tikzpicture}

}
\newsavebox\dufadx\sbox\dufadx{
\begin{tikzpicture}[
    block/.style = {
      shape = rectangle,
      rounded corners,
      draw, thick,
      minimum height = 2.1cm,
      minimum width  = 3.5cm,
      fill = c1,
    }
    ]

  \foreach \i in {-2,-1,...,2}{
    \def\dx{0.29}
    \draw[thick] (2.2,\i*\dx+0.2) -- (0,\i*\dx+0.2);
  }
  \node[block,align=center] (ufa) at (0.3,0.2) {};

  \foreach \i in {-2,-1,...,2}{
    \def\dx{0.29}
    \draw[thick] (-1.9,\i*\dx) -- (0,\i*\dx);
  }
  \foreach \i in {-2,-1,...,2}{
    \def\dx{0.29}
    \draw[thick] (1.9,\i*\dx) -- (0,\i*\dx);
  }
  \node[block,align=center] (ufa) at (0,0) 
    {$\frac{du}{dx}$ \\ $\propto$ \\
     $\frac{d\varphi}{dx}\left(\braket{\hat{\mathcal C}}^+ - \braket{\hat{\mathcal C}}^-\right)$};

\end{tikzpicture}

}
\newsavebox\result\sbox\result{\input{result.tex}}

\newcommand{\figuretext}{
  \emph{Left}: Quantum UFA fit to the training data.
  \emph{Center \& right}: Trajectories of \qmd \& DeepMoD basis function coefficients ($w_i$) during training.
}

\title{\bf Quantum Model-Discovery}
\author[1,2]{Niklas Heim}
\author[1]{Atiyo Ghosh}
\author[1,3]{Oleksandr Kyriienko}
\author[1]{Vincent E. Elfving \thanks{Corresponding author: \href{mailto:vincent.elfving@quandco.com}{vincent.elfving@quandco.com}}}
\affil[1]{Qu \& Co B.V., PO Box 75872, 1070 AW, Amsterdam, The Netherlands}
\affil[2]{Artificial Intelligence Center, Czech Technical University, Prague, CZ 120 00}
\affil[3]{Department of Physics and Astronomy, University of Exeter, Stocker Road, Exeter EX4 4QL, United Kingdom}

\date{\today}

\begin{document}
\maketitle

\begin{abstract}
  \noindent 
  Quantum computing promises to speed up some of the most challenging problems in science and engineering. 
  Quantum algorithms have been proposed showing theoretical advantages in applications ranging from chemistry to logistics optimization. 
  Many problems appearing in science and engineering can be rewritten as a set of differential equations. 
  Quantum algorithms for solving differential equations have shown a provable advantage in the fault-tolerant quantum computing regime, where deep and wide quantum circuits can be used to solve large linear systems like partial differential equations (PDEs) efficiently.
  Recently, variational approaches to solving non-linear PDEs also with near-term quantum devices were proposed. 
  One of the most promising general approaches is based on recent developments in the field of scientific machine learning for solving PDEs. We extend the applicability of near-term quantum computers to more general scientific machine learning tasks, including the discovery of differential equations from a dataset of observations.
  
  We use differentiable quantum circuits (DQCs) to solve equations parameterized by a library of operators, and perform regression on a combination of data and equations.
  Our results show a promising path to quantum model discovery (\qmd), on the interface between classical and quantum machine learning approaches. We demonstrate successful parameter inference and equation discovery using \qmd on different systems including a second-order, ordinary differential equation and a non-linear, partial differential equation.
\end{abstract}

\section{Introduction}%
\label{sec:introduction}

Recent advances in automatic differentiation in classical computers have led to novel machine learning based approaches to scientific computing, often referred to as scientific machine learning.
For example, incorporating neural networks into classical ordinary differential equation solvers \cite{Chen2018a}, or manipulating neural network derivatives towards prescribed differential equations \cite{Raissi2019}.
The applications of such techniques have included inference of unobserved states \cite{Mao2020}, inference of differential equation parameters \cite{Tartakovsky2018} and approaches to solve differential equations \cite{Raissi2019}, or discover differential equations from observated data \cite{Brunton2016a}. 
Such methods depend on machine learning techniques and are often end-to-end differentiable.

Recent progress in automatic differentiation on quantum computers \cite{Mitarai2018, vidal2018calculus, Crooks2019a, killoran2019, Mari2021, Banchi2021measuringanalytic, Cerezo2021, hubregtsen2021singlecomponent, hubregtsen2021singlecomponent, kyriienko2021} introduce the prospect of extending recent classical results in scientific machine learning to a quantum setting.
These techniques have already led to quantum methods of solving differential equations \cite{Kyriienko2020} which are analogous to methods involving classical neural networks \cite{Raissi2019}.
However, the existence of quantum automatic differentiation opens the door to further cross pollination between classical scientific machine learning and quantum computing.

In this work, we target an important task in scientific machine learning and modelling in general; while many systems can be modelled using differential equations, and their solving constitutes an important task in phenomenological understanding, in many practical settings these equations are not (fully) known. In such cases, one may have access to observations on some target system, and some initial idea of the dynamics. 

Parameter inference in differential equations is an essential part of combining theoretical models with empirical observations.
Oftentimes a given mechanism might be speculated for an observed process in the form of a differential equation, but whose coefficients must be inferred from empirical observations. 

Model discovery (often referred to as equation discovery) tries to find a mathematical expression
that describes a given dataset. We specifically tackle the problem of
discovering a differential equation only from observations of a given system.
By \emph{discovery} we mean an interpretable (i.e. human readable) model in
symbolic form, which can be converted directly into a differential equation.

This task is generally not suited for treatment by finite-differencing techniques, since such applications require many repeated simulations. Instead, to treat such a case of mixture of equation solving and model discovery efficiently, we generalize two techniques from the realm of classical scientific machine learning to a quantum setting: differential equation parameter inference and differential equation model discovery. 
We also highlight conceptual similarities between scientific machine learning and quantum computing that might point to further research avenues.

\begin{figure}[ht]
  \centering
  \resizebox{\textwidth}{!}{\input{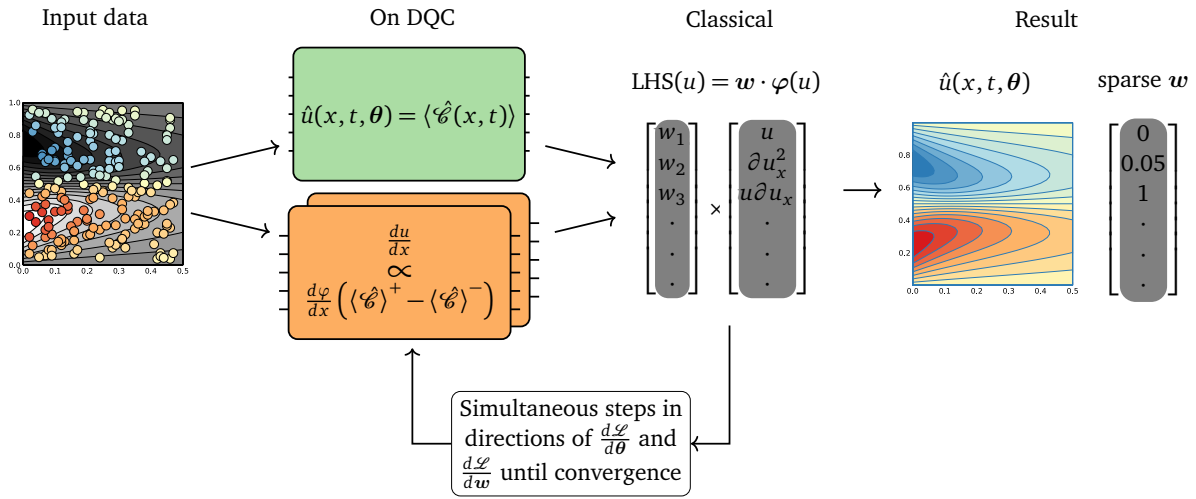}}
  \caption{Schematic of the quantum model discovery approach (\qmd). The solution to the differential equation is represented by a (quantum)
    universal function approximator (UFA) with variational angles $\bm\theta$.
    Both the forward pass and the gradients w.r.t. parameters $\bm\theta$ \emph{as well as}
    gradients w.r.t. inputs (in this example $x$ and $t$) are computed on the DQC.
    \qmd learns a sparse $\bm w$ which chooses only a few functions from the library of basis functions $\bm\varphi$.
    Gradients w.r.t. the basis function coefficients $\bm w$ are computed classically.
    The figure depicts how \qmd learns the solution $\hat u$ and a sparse $\bm w$ that only
    contains the two basis functions that are necessary to represent the input data from Burgers' equation.
    }%
  \label{fig:qmd_schematic}
\end{figure}

We first recap how quantum computers can be used to represent solutions to differential equations via variational procedures. 
Then we introduce the broad field of scientific machine learning and propose how quantum computers can be used for these nascent developments. 

\paragraph{Universal Function Approximation on Quantum Devices.}%
\label{par:universal_function_approximation_on_quantum_devices}

For the past several decades, the number of transistors on a similarly-sized integrated chip doubles roughly every two years of so; this is known as Moore's law \cite{Moore1965} and is an observation rather than a guarantee. In the past 10 years or so, a shift has started where more computational power can only be reached by parallelization rather than minimization or increasing clock speeds \cite{theis2017}; this is due to the fundamental limitations of electronics. While alternative classical hardware proposals are interesting, such as photonic \cite{krishnamoorthy2009} or biological \cite{Bonnet2013} computers, these will quickly hit a wall of unsustainable exponential speed increase too. \\

A completely different compute paradigm, quantum computing, has existed for over 4 decades but has only recently emerged as a truly promising next stage of computational hardware developments. Large corporate players including IBM \cite{Kandala2017b} and Google \cite{Arute2019a}, as well as a range of startups including IonQ, Pasqal, Xanadu and Rigetti, are investing large amounts of funding for such hardware, and with that the development of quantum computational algorithms has gotten a resurgence from the initial fundamental algorithms like Shor's \cite{Shor1995} or Grover's \cite{Grover1996} algorithms. Some of the promising areas where universal fault-tolerant quantum computing hardware could show a possible quantum advantage include quantum chemistry simulations \cite{Cao2018e, elfving2020quantum} and quantum-enhanced machine learning \cite{Biamonte}.\\

Recent algorithm developments have also focused on compatibility with current noisy intermediate scale quantum (NISQ) devices \cite{Preskill2018c}. Such algorithms are typically variational in nature, and are being applied to fields such as quantum chemistry \cite{Kandala2017b, Elfving2020b} and heuristic optimization \cite{Farhi2014, Borle2020}. \\

Furthermore, such variational quantum algorithms also offer a promising platform for the field of Machine Learning, which largely depends on training models to fit datasets and making predictions. Quantum Machine Learning (QML) algorithms have been developed for solving classification \cite{Havlicek2019SupervisedSpaces} and regression \cite{Mitarai2018} tasks; the techniques include quantum versions of kernel methods \cite{Schuld2018a, SchuldSQMLareKernel}, support vector machines \cite{Rebentrost2013, Liu2020f}, generative adversarial networks \cite{DallaireDemers2018, Lloyd2018}, Born machines \cite{coyle2020born}, Boltzmann machines \cite{Amin2016}, quantum neural networks \cite{Mitarai2018, Benedetti2019, Abbas2021} and more \cite{Biamonte}.\\

One regression task, fitting a continuous function $f(x)$ to a dataset of scalar-valued points $(x_i, y_i)$ in k-dimensional space, was analyzed for example in \cite{Mitarai2018}. A quantum neural network, consisting of a quantum feature map circuit, a variational quantum circuit and a Hamiltonian operator was used to represent a parameterizable function $f(x)$ that could be evaluated after training using gradient descent on analytically computed gradients of the loss function with respect to the variational circuit parameters.\\

In Ref. \cite{Kyriienko2020}, Differentiable Quantum Circuits (DQCs) paradigm was introduced. A DQC
implements a quantum neural network that can be analytically differentiated with respect to variational parameters \emph{and} with respect to the network inputs.
With automatic differentiation w.r.t. inputs it became possible to solve differential equations in a variational approach. In the current contribution, we will briefly summarize how this is done and we draw more parallels to classical scientific machine learning.\\

\paragraph{Physics-Informed Machine Learning}%
\label{par:physics_informed_machine_learning}

Physics informed neural networks (PINNs) comprise a a recent technique in scientific machine learning of particular relevance to this current work, since they bear strong conceptual similarities to DQCs.
We see both DQCs and PINNs as universal function approximators (UFAs) which can be differentiated with respect to their inputs. Furthermore, both can be optimized so that such derivatives can be manipulated.
Consequently, we expect that future applications of PINNs may well translate naturally to a DQC setting. Indeed, part of this current work is to demonstrate how such cross-fertilization between the fields might take place.

Introduced in \cite{Lagaris1997}, the essence of PINNs is to use (higher-order) derivatives of neural network outputs with respect to neural network inputs as terms with which to train neural networks.
Consequently, neural networks can be trained to satisfy a given differential equation.
Recently published work \cite{Raissi2019} has cause a surge in popularity of PINNs.
For example, applications of PINNs have been found in diverse fields such as aerodynamic flow modelling \cite{Mao2020} and cardiac activation mapping \cite{Costabal2020}. 
Apart from solving forward problems, they have also found use in solving inverse problems \cite{Mao2020} and optimal control \cite{Wang2021}.
See \cite{KarniadakisNature2021} for a more complete recent review of PINNs and their applications.

Due to the conceptual similarity betweens PINNs and DQCs, we anticipate ample scope for cross-pollination of ideas and techniques between these lines of research. 
Of particular relevance to this current work is the application of PINNs to equation discovery \cite{Both2021}, which we cover in more detail in subsequent sections.

\paragraph{Classical Equation Discovery}%
\label{par:classical_equation_discovery}
A popular technique in equation discovery is to overparameterize a candidate differential equation solution in terms of a library of potential terms, and then fit to data while encouraging sparsity in differential equation coefficients \cite{Brunton2016a} through an appropriate regularization term.
The resulting non-zero coefficients provide an interpretable differential equation representation of a process which fits the data.
While popularized in the context of ordinary differential equations (ODEs) \cite{Brunton2016a}, such techniques have been extended to the context of partial differential equations (PDEs) \cite{Rudy2017}.

Such sparsity-driven equation discovery schemes have recently been introduced to PINNs \cite{Both2021}. 
We are not aware of any approaches to equation discovery which levereage quantum computation.
Given the conceptual similarity between PINNs and DQCs, it is natural to extend such equation discovery techniques to a quantum setting, which is the purpose of this work.

Before we introduce our approach to \emph{quantum model discovery} (\qmd \cite{patent}) in Sec.~\ref{sub:discovering_diffeq} we briefly introduce the DQC strategy that uses \textit{quantum} neural networks as universal function approximators in Sec.~\ref{sub:quantum}.

\paragraph{Our Contributions.}%
\label{par:our_contributions}

Our main contribution is to extend DQCs to wider applications in scientific machine learning, namely:
\begin{itemize}
    \item parameter inference in differential equations
    \item discovery of unknown differential equations from empirical measurements
\end{itemize}

We demonstrate our approaches on two exemplary ordinary differential equations (ODE) and on one partial differential equation (PDE).

\section{Methods}%
\label{sec:method}
Differential equations lie at the heart of many engineering disciplines. 
However, using traditional, mesh-based methods such as finite-differencing or
finite-elements do not lend themselves well to several important applications, such as optimal control or solving inverse problems, since such applications require many repeated simulations.
Therefore, it is desirable to find cost-effective alternatives such as
data-driven differentiable surrogate models which can be optimized via gradient-based means.
On classical hardware one way of obtaining differentiable surrogates that respect physical prior 
information is via \emph{physics-informed neural networks} (PINNs).
They can be used to solve PDEs given an equation, boundary conditions, and (optionally) some measurements.
Our work builds on top of the quantum PINN analogue \citep{Kyriienko2020} which
we briefly describe in Sec.~\ref{sub:solving_diffeq}.

We then demonstrate how further scientific machine learning techniques can be extended to a quantum setting. 
In Sec.~\ref{sub:inferring_diffeq}, we introduce quantum methods for parameter inference in differential equations. Sec.~\ref{sub:discovering_diffeq} describes how to \emph{discover} equations from measurements on quantum computers. By discovering we mean finding a unknown differential equation only
from a given dataset and a predetermined (large) library of possible terms.

\subsection{Differentiable Quantum Universal Function Approximators}\label{sub:quantum} 
In DQC, quantum neural networks act as differentiable universal function approximators. These are 
quantum circuits consisting of three components: a quantum feature map, a variational ansatz, and a Hamiltonian operator for readout/encoding of the function. 
These elements act on an array of qubits and alter the system wavefunction in a specific way.

The feature map encodes the input variable $x$ using a predefined non-linearity $\varphi$
to the amplitudes of a quantum state $\fm(x)\ket{0}^n$, where $\ket{0^n}$ is some initially
prepared state. In the simplest case this can be implemented with a single layer
of rotations
\begin{equation}
  \fm(x) = \bigotimes_{j=1}^n \hat R_{\alpha,j}(\varphi_j(x)).
\end{equation}
Furthermore, as an example, choosing the non-linearity as 
\begin{equation}
  \varphi_j(x) = j \cdot \text{arccos}(x)
\end{equation}
results in a Chebyshev basis in the amplitudes of $\fm(x)\ket{0^n}$. The state now represents a basis 
of $n$ Chebyshev polynomials which can be read out by measuring for example in the Z basis on each qubit.

With a subsequent variational circuit (with parameters $\bm\theta$)
the Chebyshev basis functions are combined; in the limit of large and controlled entanglement, such a setup allows access to up to $\mathcal{O}(2^n)$ 
Chebyshev polynomials, due to the chaining and nesting properties of products of these polynomials. 
Similar effects occur for Fourier-type feature maps and other universal basis function sets. 
A popular choice for $\vc$ is the so-called hardware efficient ansatz (HEA) with parameterized rotations 
and CNOT gates between neighboring qubits on the device.

The output $\uh(x,\bm\theta)$ of the circuit is then computed with a Hamiltonian operator of choice, $\cost$,
such that the final universal approximator for a scalar input variable $x$ and variational
parameters $\bm\theta$ then reads
\begin{equation}
  \label{eq:ufa}
  \uh(x,\bm\theta) = \bra{0^n}\fm^\dagger(x)\vc^\dagger\cost\vc\fm(x)\ket{0^n}
\end{equation}
Under some conditions, equation~\ref{eq:ufa} is fully analytically differentiable via the parameter
shift rule~\citep{Mitarai2018} which requires two evaluations of the same circuit
with shifted angles for each circuit-level parameter. As a particular example, where the feature map 
consists of single-qubit rotations, we can write the derivative w.r.t. $x$ as follows
\newcommand{\fmp}{\hat{\mathcal U}_{\varphi,j,j'}^\pm(x)}
\newcommand{\fmd}{\hat{\mathcal U}_{\varphi,j,j'}^{\pm,\dagger}(x)}
\begin{align}
  \label{eq:paramshift}
  \frac{d}{dx}\uh(x,\bm\theta) &= \frac{1}{2}\frac{d\varphi}{dx} (\braket\cost^+ - \braket\cost^-),\\
  \braket\cost^\pm &= \sum_{j'} \bra{0^n} \fmd  \vc^\dagger  \cost  \vc  \fmp \ket{0^n},\\
  \fmp &=\bigotimes_{j=1}^n  R_{y,j}(\varphi(x)\pm\frac{\pi}{2}\delta_{j,j'}).
\end{align}

Please note that we so-far considered a single-dimensional independent variable $x$. However, the DQC technique can straightforwardly be extended to multiple variables $x, y, t$ etc by using multiple feature maps, one for each variable, and differentiating only those feature maps relevant to the variable differentiation of interest. An example application where this is shown in Ref. \cite{paine2021}.

Furthermore, the example shows the regular parameter shift rule, which only applies to involutory 
quantum generators in the feature map. Generalized parameter shift rules exist for arbitrary generators, and hence for arbitrary feature map circuits \cite{kyriienko2021}. In some cases, more intricate feature maps can be highly beneficial for increased expressivity of the quantum universal approximator. 

\subsection{Solving Differential Equations}%
\label{sub:solving_diffeq}
We begin by recapping how to use automatically differentiable universal function approximators (either quantum or classical) to solve partial differential equations. 
This section unifies existing treatment of using neural networks to solve differential equations \cite{Raissi2019} and analogous quantum methods \cite{Kyriienko2020}.
We write a partial differential equation in the form
\begin{equation}
  \label{eq:de}
  F(u, \ut, \ux, \uxx,\,\dots\,, x, t) = 0,
\end{equation}
where $x$ and $t$ are dependent variables and $u(x,t)$ is the solution we seek.
For simplicity we will further refer to the LHS of equation~\ref{eq:de} as
$F(u,x,t)$ omitting terms like $\ux$ or $\ut$ which in principle are operators acting on $u$.  
Let's imagine we had access to a trial function $u(x,t)$ and we would like to ask how well 
it solves the differential equation system. To find the quality of an approximate solution 
$\uh(x,t,\bm\theta)$ parameterized by $\bm\theta$ we can formulate a loss function
\begin{equation}
\label{eq:diffeqloss}
  \mathcal L(\bm\theta) = \mathcal L_F + \mathcal L_\Omega,
\end{equation}
where $\mathcal L_F$ is derived from the differential equation \ref{eq:de}
\begin{equation}
  \mathcal L_F = \frac{1}{M} \sum_{i=1}^M
                 \norm{F(\uh,x_i,t_i)}_2.
\end{equation}
and is thus minimized $F(u,x,t) = 0$, i.e. when $\uh(x,t,\bm\theta)$ satisfies the partial differential equation dynamics.
The second loss term $\mathcal L_\Omega$ includes all boundary conditions which
have to be met in order to fully specify the DE.\\

In order to effectively train such a model we ideally would like a universal approximator $\uh$
and access to its higher order derivatives both with respect to inputs $x$ and $t$,
as well as parameters $\bm\theta$. Recent advances in automatic differentiation~\citep{Gune2018}
have facilitated this using neural networks on classical hardware.  
On quantum hardware we leverage \emph{differentiable quantum circuits} (DQC)~\citep{Kyriienko2020} which implement a universal approximator based on a quantum featuremap $\fm(x)$ and a variational circuit $\vc$.

\subsection{Inferring Differential Equation Parameters}
\label{sub:inferring_diffeq}
As a first demonstration of how scientific machine learning can translate to novel quantum algorithms, we consider parameter inference in differential equations.
In many situations, an overall structure of a differential equation is known, but specific coefficients must be inferred from data. 
For example, one might parameterize a given differential equation of some scalar variable $u$ with some parameters $\bm w$ to be identified:
\begin{equation}
  \label{eq:inferdewithweights}
  F_{\bm w}(u,x,t) = 0,
\end{equation}
where $F=\bm{w}_{1} \uxx(x,t) + \bm{w}_2 x$ is one example, and the coefficients $\bm w$ must be inferred from some given data set. We denote a given set of $N$ data points with $\{(x_i, t_i, y_i): i=1,2\dots N \}$, where $x_i$ and $t_i$ denote independent variables and $y_i$ is a scalar dependent variable.   We assume that $F$ is constructed so that there is no $\bm w$ such that $F=0$ for every choice of $u$ to avoid trivial solutions.

For example, to infer a one-dimensional diffusion coefficient one might define $F$ as follows:

\begin{equation}
  F_{\bm w} = \bm w \uxx - \ut,
\end{equation}

where we have been mindful not to introduce a trainable coefficient to $\ut$, which would allow \eqref{eq:inferdewithweights} to be satisfied trivially with coefficients as zero.

To infer parameters in such settings, we introduce a loss function

\begin{equation}
  \label{eq:inferdeloss}
  \mathcal L(\bm\theta, \bm w) = \mathcal L_F + \mathcal L_d.
\end{equation}

Note that unlike \eqref{eq:diffeqloss}, we do not include a term analagous to $\mathcal L_\Omega$ to represent boundary conditions.
Although such terms could be introduced if boundary conditions are known, in principle the dynamics of the interior of the domain are inferrable from the data alone, and consequently boundary effects are not necessarily important to include.

In this loss function,

\begin{equation}
  \mathcal L_F = \frac{1}{M} \sum_{i=1}^M
                 \norm{F_{\bm w}(\uh,x_i,t_i)}_2
\end{equation}

ensures that the relevant differential equation dynamics are captured and

\begin{equation}
  \mathcal L_d = \frac{1}{N} \sum_{i=1}^N
                 \norm{\uh(x_i,t_i,\bm\theta) - y_i}_2
\end{equation}

provides incentive for the empirical dataset to be fit.

Note that we treat the loss function as a function of both the variational parameters, $\bm\theta$, as well as the inferred parameters, $\bm w$. 
In our experiments, we optimize both these parameters on an equal footing, i.e. they are trained with the same learning rate and same optimizer, but in principle a more general optimization scheme also suffices.

\subsection{Discovering Differential Equations}%
\label{sub:discovering_diffeq}

Instead of solving a \emph{known} differential equation this section describes
how to find an \emph{unknown} DE just from measurements and a library of plausible
basis functions.

As in section~\ref{sub:inferring_diffeq}, we start by introducing a parameterized differential equation which can be represented in a general form

\begin{equation}
  \label{eq:discoverdewithweights}
  F(u,x,t) = 0,
\end{equation}
where again the coefficients $\bm w$ must be inferred from some given data set. However, we parameterize $F$ such that it represents a large family of PDEs of which we expect a given instance to fit the observed data.
In particular, in the context of this work, we draw our attention to dynamical systems of the form 
\begin{equation}
  \label{eq:regprob}
  \ut = \bm w \cdot \bm\varphi(u),
\end{equation}
where $u$ is a scalar dependent variable and $\bm\varphi(u)$ is a library of
potentially important functions such as
\begin{equation}
  \label{eq:basis}
  \bm\varphi(u) = [1, x, t, u, u^2, u^3, \ux, \uxx, u\ux, \dots]^T.
\end{equation}
The library $\bm\varphi$ can be large if there is little prior knowledge about the
terms that are present in the data, or contain only few specific terms that are
likely to have generated the given dataset.
Finding a solution that is \emph{sparse} in $\bm w$ results in a simple expression
for the right-hand side (RHS) of equation~\ref{eq:regprob}. Hence, we can perform
equation discovery by finding a sparse coefficient vector $\bm w$.

Equation~\ref{eq:regprob} can be solved via STRidge regression (\cite{Brunton2016a}) if one
has access to measurements of both $u$ and $\ut$ which is rare\footnote{If
measurements of $\ut$ are not directly available they can be computed e.g.  via
the Savitzky-Golay filter (\cite{Schaferb}), but such methods typically
require very dense measurements.}.
\emph{Physics-informed neural networks} (PINN by \cite{Raissi2019}) which are normally used
for \emph{solving} DEs
can circumvent this problem by learning a surrogate $\uh$ as described in Sec.~\ref{sub:solving_diffeq}.
The surrogate is fully differentiable and therefore gives easy access to the
derivatives needed to solve equation~\ref{eq:regprob}.
Combining PINNs and equation~\ref{eq:regprob} to DeepMoD (\cite{Both2021}) results in
a fully automated framework that can discover DEs purely from data and a
pre-specified library of basis functions $\bm\varphi(u)$.

On classical hardware neural networks are used to represent the approximate solution $\uh$.
To perform equation discovery on NISQ devices we replace the classical neural
networks with quantum universal function approximators implemented on DQCs resulting
in our \emph{Quantum Model Discovery} (\qmd) approach \cite{patent}. This approach is schematically depicted in figure~\ref{fig:qmd_schematic}.

The \qmd loss is composed of three terms: the data-loss $\mathcal L_d$, the
DE (or physics) loss $\mathcal L_p$, and the regularization:
\begin{equation}
  \label{eq:discoverdeloss}
  \mathcal L(\bm\theta, \bm w) = \lambda_d \mathcal L_d + \lambda_p\mathcal L_p + \lambda_r\norm{\bm w}_1.
\end{equation}

Note, that like equation \ref{eq:inferdeloss}, we do not include a term to impose boundary conditions. 
If boundary conditions are known, such a term might be introduced. 
However, we focus on situations where the interior PDE dynamics are inferred from data alone.
Also, as in equation \ref{eq:inferdeloss}, we explicitly write the loss in terms of the variational parameters of the UFA ($\bm\theta$), and the coefficients of the learnt differential equation ($\bm w$).
We optimize over both these parameters on an equal footing, using the same optimizer and learning rate.
The term $\lambda_r\norm{\bm w}_1$ encourages sparsity in the solutions of $\bm w$ and is standard practice in statistical learning \cite{Hastie2009}.

The data-loss is the standard mean squared error (MSE)
\begin{equation}
  \label{eq:mse}
  \mathcal L_d = \frac{1}{N} \sum_{i=1}^N
                 \norm{\uh(x_i,t_i,\bm\theta) - y_i}_2,
\end{equation}
running over $N$ data points with temporal, spatial and scalar responses given by $t_i$, $x_i$ and $y_i$ respectively.
The differential equation loss is computed via the linear combination of basis functions
\begin{equation}
  \mathcal L_p = \frac{1}{M} \sum_{i=1}^M
                 \norm{\uht(x_i,t_i,\bm\theta) - \bm w \bm\varphi(\uh)}_2.
\end{equation}
The points $(x_i,t_i)$ at which the DE loss is evaluated can be chosen
independently from the data grid of equation~\ref{eq:mse} via a fixed, random, or
adaptive grid which gives this method great flexibility.

Generalizing \qmd to higher dimensional $\bm u = [u_1,u_2,\,\dots]^T$ only
requires taking care of matching dimensions (e.g. the coefficient vector $\bm
w$ turns into a matrix). In the equation discovery setting it makes sense to
prevent the surrogates for the variables $u_i$ from interaction, because we
want the interaction of terms to occur only via the library $\bm\varphi$.
Therefore, we use one individual UFA per operator instead of one large UFA 
with multiple outputs, that would allow interaction outside the library.


\section{Experiments}%
\label{sec:experiments}

In this section, we demonstrate quantum scientific machine learning to be a viable technique.
In addition to demonstrating parameter inference, we also compare our \qmd to its classical, neural network based
counter part (DeepMoD). 
We generate data from several different differential equations and demonstrate applications of \qmd.

\subsection{Damped Harmonic Oscillator}%
\label{sub:damped_harmonic_oscillator}

We start by demonstrating the potential of quantum neural networks for parameter inference and equation discovery in second-order ODES.

We parameterize a damped harmonic oscillator as follows:
\begin{equation}
  \frac{d^2 u_d}{dt_d^2} = -\omega^2 u_d -\alpha\frac{du_d}{dt_d},
\end{equation}
where $u_d$ is the position, $\omega$ the frequency, and $\alpha$ the damping coefficient
of the oscillator. The index $d$ indicates that this is the ODE before non-dimensionalization.
We non-dimensionalize the equation such that the data lies in within the range
$(-1,1)$ by defining $u_d = u u_c$ and $t_d=t t_c$ which results in 
\begin{equation}
  \label{eq:damped-harmonic-ode}
  \frac{d^2u}{dt^2} = -\omega^2 t_c^2 u -\alpha t_c^2\frac{du}{dt},
\end{equation}
With the constants $u_c=1\text{\,m}$, $t_c=2\pi\text{\,s}$, $\omega=1.5\frac{1}{\text{s}}$, and
$\alpha=0.5\frac{1}{\text{s}}$ we obtain training data as shown on the left plot of figure~\ref{fig:damped-harmonic-param-id}.

\subsubsection{Inferring Differential Equation Parameters}%
\label{ssub:parameter_identification}

As a first step we assume that the form the governing differential equation is
known and we must infer the values of the constants
$\omega$ and $\alpha$. 
We compare the method outlined in section~\ref{sub:inferring_diffeq} with an implementation of DeepMoD without regularization coefficients. 
The optimization problem in \eqref{eq:inferdeloss} with the specialized form of \eqref{eq:inferdewithweights} is defined as:
\begin{equation}\label{eq:varphi-param-id}
  F_{\bm w}(u,x,t) =  \frac{d^2u}{dt^2} + \bm w \left[t_c^2 u, t_c^2\frac{du}{dt}\right]^T.
\end{equation}
We include the known non-dimensionalization constants in the basis functions to scale the problem so that it is more amenable
to standard optimization hyperparameters (not doing this would result in the models having to identify coefficients $\gg 10$).

Figure~\ref{fig:damped-harmonic-param-id} shows the data and the obtained \qmd fit $\uh$
on the left. We use a quantum neural network with 5 qubits as the UFA to approximate $\uh$. We
reach an MSE loss value of $\mathcal L_d=10^{-7}$ after 200 BFGS optimization steps.  A
neural network with one hidden layer of size 32 and $\tanh$ activations reaches
similar MSE.
The central and right plot of figure~\ref{fig:damped-harmonic-param-id} show the
trajectories of the coefficients of the two basis functions
(equation~\ref{eq:varphi-param-id}), namely $\omega$ and $\alpha$, during training.
The target values for the two coefficients are $\omega=1.5$ and $\alpha=0.5$.

In this experiment we found good solutions using $\lambda_d=1$ (dataloss),
$\lambda_p=10^{-5}$ (physics loss) for both \qmd and DeepMoD within \eqref{eq:inferdeloss}.
The identified parameter values are shown in table~\ref{tab:damped_harmonic_param_id}.
\begin{figure}
  \centering
  \resizebox{\textwidth}{!}{\input{damped-harmonic-param-id.tikz}}
  \caption{Damped harmonic parameter \textit{identification}. \figuretext
  Both \qmd and the classical DeepMoD infer the correct target parameters
  $\omega^2=2.25$ and $\alpha=0.5$.
  }%
  \label{fig:damped-harmonic-param-id}
\end{figure}
\begin{table}
  \label{tab:damped_harmonic_param_id}
  \centering
  \begin{tabular}{llll}
    \toprule
             &       & \multicolumn{2}{c}{Basis function} \\
    \cmidrule{3-4}
    Equation & Model & $u$  & $\frac{du}{dt}$ \\
    \midrule
    \multirow{3}{*}{$\frac{d^2u}{dt^2} = -\omega^2 t_c^2 u -\alpha t_c^2\frac{du}{dt}$}
    & \qmd      & $\omega=\sqrt{2.249} = 1.499$               & $\alpha=0.500$ \\
    & DeepMoD   & $\omega=\sqrt{2.247} = 1.498$               & $\alpha=0.500$ \\
    & Truth     & $\omega=1.5$                                & $\alpha=0.5$ \\
    \bottomrule
  \end{tabular}
  \caption{Final basis function parameters $\omega$ and $\alpha$ as inferred by \qmd
    and DeepMoD compared to the true coefficients that generated the data (called
    \emph{truth}).
    Non-dimensionalization coefficients in the basis functions are omitted for clarity.}
\end{table}

\subsubsection{Equation Discovery}%
\label{ssub:equation_discovery}

To go a step beyond the task of parameter inference, we next aim for full "equation discovery"; we
assume that only the LHS of equation~\ref{eq:damped-harmonic-ode} (i.e. $\partial u_t^2$) is known and choose
a set of plausible basis functions $\bm\varphi$ consisting of polynomials up to
degree 4 and the first derivative of $u$ for use in \eqref{eq:regprob}.
\begin{equation}\label{eq:varphioscillator}
  \bm\varphi(u) = \left[t_c^2\, u(t),\quad
                        u_c t_c^2\, u^2(t),\quad
                        u_c^2t_c^2\, u^3(t),\quad
                        u_c^3t_c^2\, u^4(t),\quad
                        t_c \frac{du}{dt}
                  \right]^T
\end{equation}
\begin{figure}
  \centering
  \resizebox{\textwidth}{!}{\input{damped-harmonic.tikz}}
  \caption{Damped harmonic oscillator \emph{discovery}. \figuretext
    The difference to
    figure~\ref{fig:damped-harmonic-param-id} is the larger set of basis
    functions.  Both \qmd and DeepMoD correctly identify the relevant basis
    functions ($u$ and $\frac{du}{dt}$) whereas the irrelevant basis function
    coefficients are all close to zero.
  }%
  \label{fig:damped-harmonic}
\end{figure}
\newcommand{\sconst}{$<10^{-3}$}
\begin{table}
  \centering
  \begin{tabular}{lllcccl}
    \toprule
             &       & \multicolumn{5}{c}{Basis function} \\
    \cmidrule{3-7}
    Equation & Model & $u$ & $u^2$ & $u^3$ & $u^4$ & $\frac{du}{dt}$ \\
    \midrule
    \multirow{3}{*}{$\frac{d^2u}{dt^2} = -\omega^2 t_c^2 u -\alpha t_c^2\frac{du}{dt}$}
    & \qmd    & $\omega=\sqrt{2.239} = 1.496$ &\sconst&\sconst&\sconst& $\alpha=0.497$ \\
    & DeepMoD & $\omega=\sqrt{2.242} = 1.498$ &\sconst&\sconst&\sconst& $\alpha=0.496$ \\
    & Truth   & $\omega= 1.5$ &$0$&$0$&$0$& $\alpha=0.5$ \\
    \bottomrule
  \end{tabular}
  \label{tab:damped_harmonic}
  \caption{Final basis function parameters resulting from \qmd/DeepMoD training with
    a larger library of basis functions (equation~\ref{eq:varphioscillator}).
    All irrelevant coefficients are very close to zero while the inferred $\omega$
    and $\alpha$ are close to their true values.
    Non-dimensionalization coefficients in the basis functions are omitted for clarity.
  }
\end{table}
We optimize \eqref{eq:discoverdeloss} and achieve good results with a regularization parameter $\lambda_r=10^{-4}$.
Otherwise we use the same training and model setup (5 qubits, 200 BFGS steps,
$\lambda_p=10^{-5}$) as in section~\ref{ssub:parameter_identification}.
Both our quantum equation discovery \qmd and DeepMoD successfully identify the
relevant basis functions and their correct coefficients. All other basis function
coefficients are close to zero after training.

\subsection{Lotka-Volterra System}%
\label{sub:lotka_volterra_system}
We now demonstrate that coupled non-linear differential equations can be discovered with quantum circuits.
Our model system is given by the Lotka-Volterra (LV) equations:
\begin{align}
  \label{eq:lotka-volterra}
  \frac{dx_d}{dt_d} &= \alpha x_d - \beta x_dy_d\\
  \frac{dy_d}{dt_d} &= \delta x_dy_d -  \gamma y_d
\end{align}
that describe a predator-prey system where $x_d$ is the number of prey and
$y_d$ the number of predators.  The parameters $\alpha$, $\beta$, $\gamma$, and
$\delta$ define the interaction of the two species which we chose to
$\alpha=1.5$, $\beta=1$, $\gamma=1$, $\delta=1$.
Again, the index $d$ indicates dimensional quantities. We non-dimensionalize
the LV system with critical time $t_c=8$ and a critical number of individuals
$x_c=2.5$ which results in
\begin{align}
  \label{eq:lotka-volterra-nd}
  \frac{dx}{dt} &= \alpha t_c x - t_c x_c \beta xy\\
  \frac{dy}{dt} &= t_c x_c \delta xy -  t_c\gamma y.
\end{align}
The non-dimensionalization coefficients are again included in the basis
functions $\bm\varphi$ which consist of polynomials up to degree 2
\begin{equation}
  \bm\varphi(x,y) = \left[t_c x,\, t_c y,\, t_c x_c x^2 ,\, t_c x_c y^2 ,\, t_c x_c xy\right]^T.
\end{equation}

\qmd is trained with a time-series of 30 datapoints for each variable $x$ and
$y$.  For this task we need a slightly larger featuremap with 7 qubits to fit the data.
Each variable of the LV system is modeled by a separate UFA driving its own set of (identical) library functions.  
After 200 BFGS steps we discover the correct basis
function coefficients as shown in figure~\ref{fig:lotka-volterra} and
table~\ref{tab:lotka-volterra}. The \qmd loss coefficients are $\lambda_d=1$,
$\lambda_p=10^{-3}$, and $\lambda_r=10^{-5}$.

We compare \qmd to classical DeepMoD with two neural networks (one for each
variable $x$ and $y$) with one hidden layer of size 32.
Classical DeepMoD receives the same data. For this dataset we observe that the
classical neural networks suffer from stronger gradient pathologies which could
only be resolved by decreasing the physics loss coefficient to $\lambda_p=10^{-6}$
which weakens the training signal on the basis function coefficients.
This explains why DeepMoD needs more training steps to converge (see figure~\ref{fig:lotka-volterra}).
\begin{figure}
  \centering
  \resizebox{\textwidth}{!}{\input{lotka-volterra.tikz}}
  \caption{Lotka-Volterra discovery. \figuretext
  Both \qmd and DeepMoD
  correctly identify the relevant basis functions for each variable.}%
  \label{fig:lotka-volterra}
\end{figure}
\begin{table}
  \label{tab:lotka-volterra}
  \centering
  \begin{tabular}{llccccc}
    \toprule
             &       & \multicolumn{5}{c}{Basis function} \\
    \cmidrule{3-7}
    Equation & Model & $x$  & $y$ & $x^2$ & $y^2$ & $xy$ \\
    \midrule
    \multirow{3}{*}{$\frac{dx}{dt} = \alpha t_c x - t_c x_c \beta xy$}
    & \qmd      & \caleft{$\alpha = 1.483$}    & $0.001$           & $-0.001$   & $-0.007$   & \caleft{$\beta=0.986$} \\
    & DeepMoD   & \caleft{$\alpha = 1.475$}    & $<10^{-3}$        & $<10^{-3}$ & $-0.003$   & \caleft{$\beta=0.979$} \\
    & Truth     & \caleft{$\alpha = 1.5$}      & $0$               & $0$        & $0$        & \caleft{$\beta=1$}\\
    \midrule
    \multirow{3}{*}{$\frac{dy}{dt} = t_c x_c \delta xy -  t_c\gamma y$}
    & \qmd      & $0.006$             & \caleft{$\delta = 0.993$}  & $0.008$    & $<10^{-3}$ & \caleft{$\gamma=0.991$} \\
    & DeepMoD   & $<10^{-3}$          & \caleft{$\delta = 0.987$}  & $0.011$    & $<10^{-3}$ & \caleft{$\gamma=0.979$} \\
    & Truth     & $0$                 & \caleft{$\delta = 1$}      & $0$        & $0$        & \caleft{$\gamma=1$}\\
    \bottomrule
  \end{tabular}
  \caption{Final basis function parameters of the LV system discovered by \qmd and DeepMoD
    compared to the true coefficients that generated the data (called
    \emph{truth}).  Non-dimensionalization coefficients in the basis functions
    are omitted for clarity. Both models identify the correct basis functions as relevant.}
\end{table}

\newpage
\subsection{Burgers' Equation}%
\label{sub:burgers_equation}

\begin{figure}
  \centering
  \includegraphics[width=\linewidth]{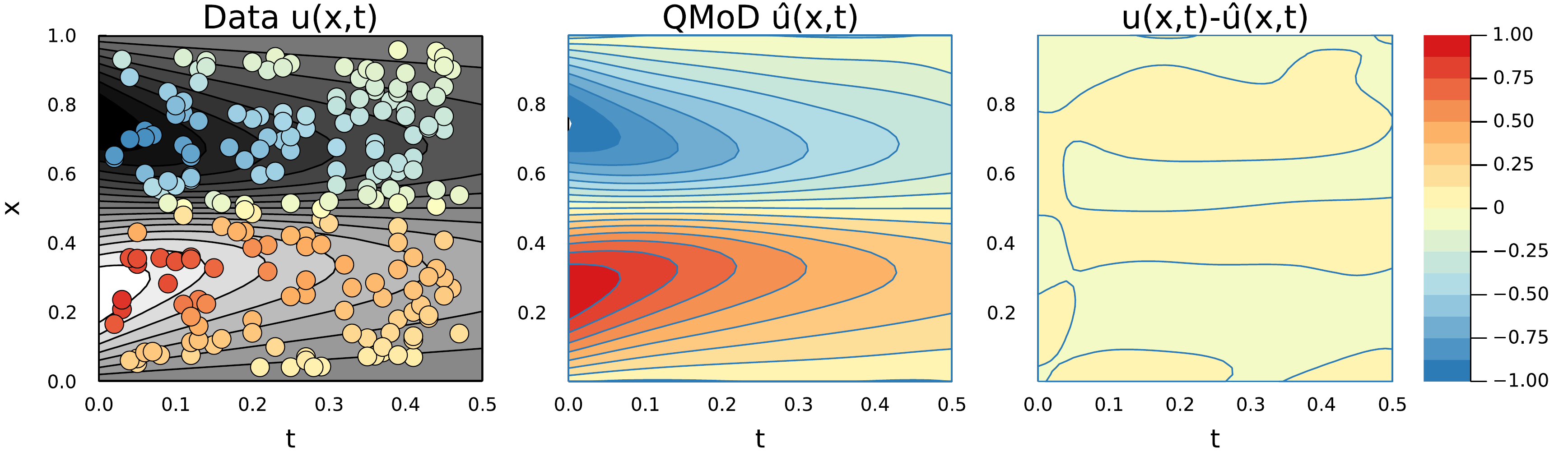}
  \caption{\emph{Left}: Scatter plot of Burgers' equation dataset and the full solution as a gray-scale contour underneath
  (parameters $\varepsilon=0.05$, $\alpha=1$, and initial condition $u(x,0)=\sin(2\pi x)$).
  \emph{Center}: Learned \qmd solution. \emph{Right}: Error of \qmd solution.}%
  \label{fig:burgers-heatmap}
\end{figure}

In this last section we present the discovery of Burgers' equation,
a non-linear second-order partial differential equation.
We chose the general (or \emph{viscous}) Burgers equation with one spatial
dimension $x$ and one time dimension $t$:
\begin{equation}
  \frac{\partial u}{\partial t} = \varepsilon\frac{\partial^2u}{\partial x^2} - \alpha u\frac{\partial u}{\partial x},
\end{equation}
where $u(x,t)$ is the velocity field, $\varepsilon$ the viscosity, and $\alpha$
the coefficient of the advection term. To obtain training data we solve
Burgers' equation on the domain $(x \times t)\in([0,1]\times[0,\frac{1}{2}])$
with parameters $\varepsilon=0.05$ and $\alpha=1$. The true solution with sinusoidal
initial conditions
\begin{equation}
  u(x,0) = \sin(2\pi x)
\end{equation}
is shown in the left plot of figure~\ref{fig:burgers-heatmap} as the gray-scale contour. We train on 200
uniformly sampled points\footnote{Note that no intricate grid definition is
needed for DQC/\qmd to solve PDEs} which are scattered on top of the left plot of figure~\ref{fig:burgers-heatmap}. For Burgers' equation we do not need to
perform non-dimensionalization as the domain is already within $(-1,1)$.
We assume that the LHS $\partial u_t$ is known and want to discover $\varepsilon$
and $\alpha$ from the library
\begin{equation}
  \bm\varphi(u) = \left[u,\quad u^2,\quad \frac{\partial u}{\partial x},\quad
                        \frac{\partial^2u}{\partial x^2},\quad
                        u\frac{\partial u}{\partial x},\quad
                        u^2\frac{\partial u}{\partial x}
                  \right]^T.
\end{equation}
To solve this PDE we need a UFA with two inputs $(x,t)$ and one output.
Therefore, we employ two input featuremaps with 6 qubits each. The two featuremaps
are stacked in parallel onto one circuit, and connected to a variational circuit with a base on all 12 qubits, effectively entangling the two feature registers.
The physics loss coefficient is set to $\lambda_p=10^{-2}$ and the
regularization to $\lambda_r=10^{-3}$.

Figure~\ref{fig:burgers-discovery} and table~\ref{tab:burgers} show that
$\varepsilon$ is discovered very precisely but the advection term $\alpha$ is
slightly too low in both models.
However, we can still conclude that \qmd and DeepMoD correctly identify the two
relevant basis functions for advection and diffusion.
To get an even closer fit to the true basis function parameters it is possible to re-run
the model only with the two relevant basis functions an switch off the regularization (as we demonstrated
in section~\ref{ssub:parameter_identification}).
\begin{figure}
  \centering
  \resizebox{\textwidth}{!}{\input{burgers.tikz}}
  \caption{Burgers' equation discovery. \figuretext The trajectory of viscosity
    $\varepsilon$ (red line) is scaled by $\frac{1}{\varepsilon}$ for better
    visibility. Both \qmd and DeepMoD discover $\varepsilon$ with good
    precision. The advection term $\alpha$ (blue line) is clearly discovered as a relevant
    term but does not match the true value $\alpha=1$ perfectly.
  }%
  \label{fig:burgers-discovery}
\end{figure}
\begin{table}
  \centering
  \begin{tabular}{llcccllc}
    \toprule
             &       & \multicolumn{6}{c}{Basis function} \\
    \cmidrule{3-8}
    Equation & Model & $u$ & $u^2$ & $\frac{du}{dx}$ & $\frac{d^2u}{dx^2}$ & $u\frac{du}{dx}$ & $u^2\frac{du}{dx}$ \\
    \midrule
    \multirow{3}{*}{$\frac{\partial u}{\partial t} = \varepsilon\frac{\partial^2u}{\partial x^2} - \alpha u\frac{\partial u}{\partial x}$}
    & \qmd    & \sconst & \sconst & \sconst & $\varepsilon=0.050$ & $\alpha = 0.838$ & \sconst \\
    & DeepMoD & \sconst & \sconst & \sconst & $\varepsilon=0.049$ & $\alpha = 0.924$ & \sconst \\
    & Truth   & $0$     & $0$     & $0$     & $\varepsilon=0.05$  & $\alpha = 1$     & \sconst \\
    \bottomrule
  \end{tabular}
  \label{tab:burgers}
  \caption{Final basis function parameters for Burgers' equation discovered by \qmd
    and DeepMoD compared to the true coefficients that generated the data. Both \qmd and DeepMoD correctly identify that
    only two basis functions are needed to describe Burgers' equation. The parameters $\varepsilon$ is found very precisely
    by both models. DeepMoD is closer to the true parameter $\alpha$ but still slightly too low.}
\end{table}

\section{Discussion \& Conclusion}%
\label{sec:conclusion}
We demonstrated that techniques from scientific machine learning can be placed in a quantum computational setting.
In particular, our numerical experiments show that relatively small quantum circuits can be used for meaningful computations in scientific machine learning.
We have demonstrated how to implement an automated parameter inference and equation discovery
framework (\qmd) using a differentiable quantum circuit strategy, and reached results that are on
par with the classical machine learning method DeepMoD.
With this demonstration of what can be achieved with DQCs the fields of quantum
computing and scientific machine learning are moving closer together.

We should note that all experiments described here involve classical simulators of quantum computers with unrealistic assumptions such as zero sampling noise, no environment noise and no measurement noise. Although many error mitigation strategies for NISQ hardware exist, it should still be investigated how well the strategies described work in more realistic settings and importantly on real quantum hardware, and which quantum hardware is most compatible/natural for this strategy.
Furthermore, due to the intrinsic scaling issues of simulating quantum many-body systems, we did not consider more than 12 qubits in the largest simulations described here, and thus also compared with only small-sized classical neural networks. It will be important to investigate the performance for larger number of qubits on quantum hardware, and compare them to realistic classical PINN architectures. 

The strategies described above work for systems governed by sets of Partial Differential Equations, but also for those governed by stochastic processes modelled as Stochastic (Partial) Differential Equations (SDEs), through the perspective of \textit{quantile mechanics} combined with DQCs as described in the recent paper on Quantum Quantile Mechanics (QQM) \cite{paine2021}. In such setting, an SDE may be learned (`discovered') based on just statistical data. This could prove valuable in the generative modelling case, where synthetic data generation of time-series is learned based on a combination of data and a parameterized SDE with a library of candidate model operators. In such real-world setting as financial markets, there is often not exact knowledge of the underlying model dynamics but there is access to historical data, where \qmd could infer the SDE from data and make better predictions compared to a purely data-only approach.

An interesting area of potential yet unexplored, is the identification of quantum computational speedup for scientific machine learning in other aspects than Neural Network enhancement, for example by combining \qmd with combinatorial optimization. 

\textit{Ethics declaration.} A patent application for the method described in this manuscript has been submitted by Qu\&Co.

\printbibliography

\end{document}